\documentclass[11pt]{article}
\usepackage[utf8]{inputenc}
\usepackage{geometry}
 \usepackage{graphicx}
 \usepackage{subfig}
  \usepackage{amsmath} 
  \graphicspath{{images33/}}
  \usepackage[usenames, dvipsnames]{color}
\usepackage{wrapfig}
\usepackage{boxedminipage}
\usepackage{cancel}
\usepackage{enumerate,amsmath,amssymb,hhline,multirow,array}
\newcommand{\RNum}[1]{\uppercase\expandafter{\romannumeral #1\relax}}
\usepackage{tikz}
\usepackage{romannum}
\usepackage{array}
\newcolumntype{M}[1]{>{\centering\arraybackslash}m{#1}}

\tikzset{
  threept/.style={
    circle,
    draw,
    inner sep=2pt,
  },
  twopt/.style={
    circle,
    draw,
    fill=black,
    inner sep=1pt,
    minimum size=1pt
  },
  cross/.style={
    cross out,
    draw=black, 
    minimum size=7pt, 
    inner sep=0pt,
    outer sep=0pt
  },
  scalar/.style={
    thick,
    dashed,
    postaction={
      decorate,
      decoration={
        markings,
        mark=at position 0.5 with {\arrow{>}}
      }
    }
  },
  spinning/.style={
    thick,
    postaction={
      decorate,
      decoration={
        markings,
        mark=at position 0.5 with {\arrow{>}}
      }
    }
  },
  spinning no arrow/.style={
    thick,
  },
  finite with arrow/.style={
    decoration={
      snake,
      amplitude=1pt,
      segment length=6pt,
      post length=2pt
    },
    decorate,
    thick,->
  },
  finite/.style={
    decoration={
      snake,
      amplitude=1pt,
      segment length=6pt,
    },
    decorate,
    thick
  }
}

\usepackage{fullpage}
\usepackage{amsmath}
\usepackage{amssymb}
\usepackage{graphicx}
\usepackage{mathrsfs}
\usepackage{wrapfig}
\usepackage{boxedminipage}
\usepackage{epsfig}
\usepackage{setspace}
\usepackage{float}
\usepackage{hyperref}
\usepackage{enumerate}
\usepackage{cite}
\usepackage[font=footnotesize,labelfont=rm]{caption}

\usepackage[numbers,sort&compress]{natbib}

\def\be{\begin{equation}}
\def\ee{\end{equation}}
\def\bea{\begin{eqnarray}}
\def\eea{\end{eqnarray}}

\numberwithin{equation}{section}

 \newcommand{\RN}[1]{%
   \textup{\uppercase\expandafter{\romannumeral#1}}%
 }
\begin{document}

\pagenumbering{arabic}

\vskip 2cm 

\let\endtitlepage\relax

\begin{titlepage}
\begin{center}
\renewcommand{\baselinestretch}{1.5}  

\vspace*{-0.5cm}

{\fontsize{19pt}{22pt}\bf{A note on size-momentum correspondence and chaos}}

\vspace{9mm}
\renewcommand{\baselinestretch}{1}  

\centerline{\large{Sandip Mahish$^{\dagger}$}\footnote{sm19@iitbbs.ac.in}, \large{Shrohan Mohapatra$^{\ddagger}$}\footnote{sm32@iitbbs.ac.in},\large{Karunava Sil$^{\dagger}$}\footnote{ks45@iitbbs.ac.in} and \large{Chandrasekhar Bhamidipati$^\dagger$}\footnote{chandrasekhar@iitbbs.ac.in}}

\vspace{5mm}
\normalsize
\textit{$^\dagger$School of Basic Sciences, Indian Institute of Technology Bhubaneswar, Odisha, 752050, India}\\
\textit{$^\ddagger$Department of Physics, University of Massachusetts, Amherst, MA 01003, US}\\
\vspace{5mm}

\begin{abstract}

The aim of this note is to explore Susskind's proposal~\cite{Susskind1} on the connection between
operator size in chaotic theories and the bulk momentum of a particle falling into black holes (see also~\cite{Susskind2,Ageev:2018msv,Susskind:2019ddc,Barbon:2019tuq,Susskind:2020gnl} for more recent generalizations), in a broad class of models involving Gauss-Bonnet(GB) and Lifshitz-Hyperscaling violating theories in AdS. For Gauss-Bonnet black holes, the operator size is seen to be suppressed as the coupling constant $\lambda$ is increased. For the Lifshitz-hyperscaling violating theories characterised by the parameters $z$ and $\theta$, the operator size is higher as compared to case $z=1,\theta=0$ (Reissner-Nordstrom AdS black holes). In the case of operators with global charge corresponding to charged particles falling into black holes, suppression of chaos is seen in general theories of gravity, in conformity with the original proposal~\cite{Susskind1} and earlier findings~\cite{Ageev:2018msv}.

\end{abstract}
\end{center}
\vspace*{0cm}


\end{titlepage}
\vspace*{0cm}

\newpage
\setcounter{footnote}{0}
\noindent

\baselineskip 15pt

\section{Introduction}

The connection between chaotic systems and gravity has received continuous attention with the emergence of new concepts such as scrambling, complexity proposals, butterfly effect etc.,~\cite{Susskind1}-\cite{Size1}. There have also been an important observations regarding a possible close relationship between gravity and quantum mechanics, such as the GR$=$QM proposal~\cite{Susskind:2017ney}. In this direction, recently, a nontrivial relation between the increase of radial momentum of a particle falling into the bulk geometry under the influence of a gravitating mass and the growth of the corresponding operator on the boundary theory was conjectured in \cite{Susskind1}. The growth of operator size in chaotic systems is found to be exponential. This exponential growth in the operator size and growth of the bulk momentum of particle was studied for charged black holes as well as empty AdS spacetimes~\cite{Susskind1}(see also~\cite{Ageev:2018msv,Susskind2}). There are further connections of above quantities with holographic complexity too~\cite{Susskind:2019ddc,Barbon:2019tuq,Susskind:2020gnl,Barbon:2020olv,Barbon:2020uux}. The size-momentum correspondence can be understood as follows. In the context of holography, considering an operator $W(t)$ on the boundary (where $t$ is the boundary time), can be thought of as injecting a neutral particle of mass $m$ in the bulk. The time evolving $W(t)$ can be written in an expansion as
\begin{equation}
W(t)=\sum_{s}W_s(t) \, ,
\end{equation}
where $W_s(t)$ is a product of a number of simple operators $\psi$, given as
\begin{equation} \label{ws}
W_s(t)=\sum_{a_1<...<a_s} c_{a_1...a_s}(t)\psi_{a_1}\cdots \psi_{a_s} \, .
\end{equation}
One can think of $c_{a_1...a_s}(t)$ as the coefficients that describe the quantum wave function of the evolving operator. The number of simple operators that form a given product as in eqn. (\ref{ws}) is called the operator size $s$~\cite{Size1,Shenker:2014cwa}.  The probability distribution for the operator size is simply
\begin{equation} \label{dist}
{\mathcal P}_s(t) = \sum_{a_1<...<a_s} \left| c_{a_1...a_s}(t) \right|^2 \, .
\end{equation}
In the bulk, as the particle evolves away from the boundary with momentum $p(t)$, and nears the horizon, the late time behavior is
\begin{equation} 
p(t) \sim e^{\frac{2\pi}{\beta}t} \, ,
\end{equation}
where  $\beta$ is the inverse temperature and related to the lapse function of black hole. Correspondingly, its dual boundary operator evolves in time too with the distribution in eqn. (\ref{dist}) shifting towards operators with larger size, becoming increasingly complex.  A further proposal then is that, the bulk momentum of a particle falling in a black hole is also proportional to the growth rate of complexity of the precursor operator $W(t)$ that created it (see e.g.,~\cite{Size1,Barbon:2019tuq}). We thus arrive at the triple-correspondence proposal of~\cite{Susskind1,Susskind:2019ddc,Susskind2}, which is also describable as the momentum-volume-complexity (PVC) relation~\cite{Barbon:2019tuq,Barbon:2020olv,Barbon:2020uux} as:
\begin{equation} \label{triple}
{\rm operator~size} \sim P(t) \sim \frac{d {\mathcal C}}{dt}
\end{equation}
where $P(t) = \beta p(t)$ is the RG corrected particle momentum~\cite{Susskind2}, and ${\mathcal C}$ is the complexity of operator $W(t)$. As discussed in~\cite{Barbon:2019tuq}, the complexity of the operator can be defined in terms of the complexity of the evolved state together with an appropriate CA\cite{Brown:2015bva,Brown:2015lvg,Lehner:2016vdi} or CV~\cite{Stanford:2014jda,Alishahiha:2015rta} prescription. Recently, in \cite{Susskind:2020gnl} further connections were made for gravitational backgrounds with dimension more than two by considering a toy model, namely the gluon splitting model, where a highly energetic gluon splits into a number of low energy gluons in the RG flow from UV to IR. In particular, the authors in \cite{Susskind:2020gnl} considered a conformal field theory on the boundary and perturbed it by smearing a gauge invariant local operator over space such that in the bulk picture the corresponding effect is to inject a finite energy $E$. For a fixed energy $E$, as the number of gluons increases due to the splitting, the energy of each gluon decreases and the theory flows from high to low energy regime. In other words, as the theory flows from UV to IR there is an increase in the number of gluons which corresponds to the growth of operator size on the boundary. 
Thus, the above correspondences are proposed to hold even if there is no black hole and one has only empty AdS space-time~\cite{Susskind:2020gnl}.\\

\noindent
It is interesting to test whether the above conjectures relating size-momentum and complexity hold for general theories of gravity, involving either higher derivative terms or other models. One expects the above conjectures to hold, as for instance, the complexity=momentum conjecture was originally proposed to setup a bridge connecting the theory of gravity and quantum mechanics, and is very general. Based on a series of interesting papers in recent times, it is evident that the above proposal will hold equally well for any classical theory of gravity including the one with higher derivative corrections. In this regards, the authors in~\cite{1,2} argued that for holographic theories the first law of complexity can be equivalently realized from the linearised Einstein’s equation involving bulk perturbation about not only the vacuum AdS but also a more general gravitational backgrounds which includes the theory of higher derivative gravity. However, for the above equivalence between the first law of complexity and the classical theory gravity to hold one must assume the validity of the CV proposal for the same theory. Gauss-Bonnet gravity is an important example when testing holographic complexity conjectures, together with Lifshitz hyperscaling violating theories. In~\cite{An:2018dbz}, the late time growth of the complexity for the Gauss-Bonnet gravity was shown to obey the Lloyd’s bound using the CV conjecture which successfully validates the application of the CV proposal in this case. On the other hand in~\cite{Barbon:2020olv} the authors has shown that the Complexity=Momentum proposal is implicitly connected to the CV conjecture as a result of the constraint on momentum of any in falling matter under gravity. Thus, being part of a larger program connecting gravity to quantum information, there are enough evidences to believe that the Complexity=momentum and size proposal may be valid in general and ideas in~\cite{Susskind:2020gnl} can be extended to perform computations in any theory of classical gravity. \\

\noindent
The aim of this paper is to conduct a similar kind of analysis as in~\cite{Susskind1,Susskind:2019ddc,Susskind2,Ageev:2018msv}, to verify the first two relations in equation-(\ref{triple}), in very  different class of gravity theories. In section-(\ref{momentum}) we study the late time behaviour of radial momentum of both neutral and charged particles falling into the charged black holes in Gauss-Bonnet and Lifshitz-Hyperscaling violating theories, as the parameters of the respective models are varied. Section-(\ref{size}) is devoted to studying the operator growth on the boundary for the models considered in section-(\ref{momentum}), using the recent proposals in~\cite{Susskind:2020gnl}. We end with remarks in section-(\ref{conclude}).

\section{Momentum of a particle falling into the black hole }\label{momentum}

For computing the bulk momentum of a particle falling into the black hole, we closely follow the analysis for charged particles in \cite{Ageev:2018msv}. In\cite{Ageev:2018msv} it was shown that depending on the value of particle charge $q$, which is greater than or less than a critical value $q_{\rm crit}$, the particle momentum behaves differently. For $q<q_{\rm crit}$, the late time growth of momentum does not depend on the value of $q$ chosen, with $q \approx q_{\rm crit}$ regime signalling scrambling time growth. When $q = q_{\rm crit}$, the particle momentum growth ceases after a certain time.  For $q>q_{\rm crit}$, the particle momentum shows an oscillating behaviour pointing towards localisation. In the following two subsections-(\ref{GB}) and (\ref{HV}),  we present our results on time evolution of momentum of both charged and neutral particles in charged Gauss-Bonnet and Lifshitz-hyperscaling violating theories, respectively by numerically solving the relevant equation of motion in each case. For the charged particles (we restrict to the case $q<q_{\rm crit}$)  the claims in~\cite{Ageev:2018msv} regarding suppression of chaos due to charge are tested in the above general theories of gravity. A further aim is to see the variation of momentum of the neutral particle at late times with the parameters of the model, and compare it with the behaviour of size studied in section-(\ref{size}).


\subsection{Charged Gauss-Bonnet AdS Black Holes }\label{GB}


Let us start by considering a planar Gauss-Bonnet AdS black hole in five dimensions given by the metric~\cite{Cai:2001dz,Cai:2009zn}:
 \bea
 && ds^2=\frac{1}{z^{2}}\left(-\phi(z)N^2dt^2+\frac{dz^2}{\phi(z)}+\frac{1}{\ell^2}d\bar{x}^2 \right),\\
 && A=\mu \left(1-\frac{z^2}{z_+^2} \right)dt,\eea
where $A$ is the gauge field and the lapse function $\phi(z)$ is:
 \be \phi(z)=\frac{1}{2\lambda \ell^2}
 \Big[1-\sqrt{1-4\lambda
\big(1-\frac{z^2}{z_+^2}\big)\big(1-\frac{z^2}{z_-^2}\big)\big(1+z^{2}(\frac{1}{z_+^2}+\frac{1}{z_-^2})\big)}\Big] \, .\nonumber 
\ee
Here, $\lambda$ is the Gauss-Bonnet coupling constant, $l$ is the AdS length scale, $z_+$ is outer horizon radius and  $\mu$ is the chemical potential in the dual quantum field theory living on the boundary $z=0$, given together with $N$ below as:
 \begin{equation}
 \mu=\frac{\sqrt{6}qN z_+^2}{2\kappa\ell}  \, ,\qquad \quad N^2=\frac{1}{2}\Big(1+\sqrt{1-4\lambda}\Big) \, .
 \end{equation}
 We can write other parameters such as mass ${\mathcal M}$, temperature $T$ and charge $Q$, in terms of chemical potential as:
\begin{eqnarray}
&&{\mathcal M}=\frac{1}{z_+^4}+\frac{2\kappa^2\mu^2\ell^2}{3N^2 z_+^2} \, , \nonumber\\
&&Q=\frac{2\mu\kappa\ell}{\sqrt{6}Nz_+^2} \, ,\nonumber\\
&&T=\frac{N}{\pi\ell^2z_+}\left(1-\frac{\kappa^2\ell^2z_+^2\mu^2}{3N^2}\right)~~~\text{where}~~\kappa=\sqrt{8\pi} \, .
\end{eqnarray}
In the extremal limit where the temperature vanishes, the outer and inner horizon will coincide, giving $z_+=z_-=z_h$. This condition results in an extremal value of parameters given as:
\begin{equation} \label{muext}
Q_{extremal}=\frac{\sqrt{2}}{z_h^3},~~\mu_{extremal}=\frac{\sqrt{3}N}{\kappa\ell z_h} \, .
\end{equation}
Now, the action of a charged particle falling in to the Gauss-Bonnet black hole is:
\begin{equation}
S=-m\int\sqrt{\frac{\phi(z)N^2}{z^2}-\frac{\dot{z}^{2}}{z^2\phi(z)}}dt+q\mu \left(1-\frac{z^2}{z_{+}^{2}} \right)dt \, .
\label{action}
\end{equation}
\begin{figure}
	\begin{center}
		{\centering
			{\includegraphics[width=5cm,height=5cm]{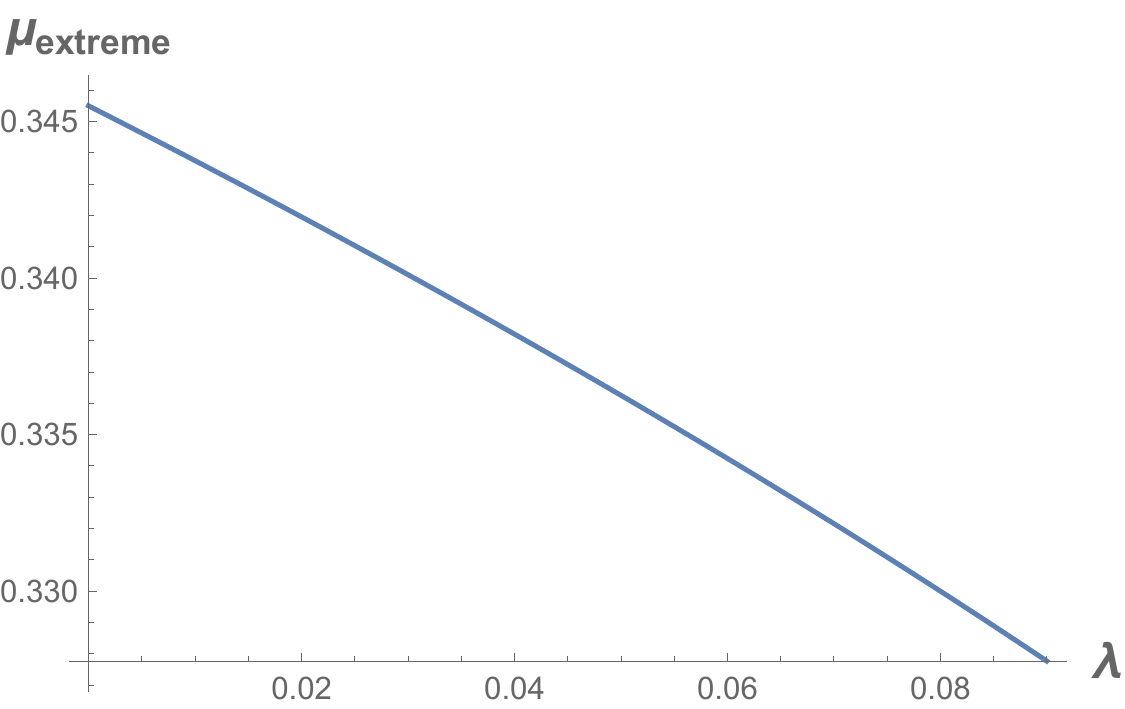} } 
			\caption{Variation of $\mu_{extremal}$ with $\lambda$ for GB }  \label{muextremal}
		}
	\end{center}	
\end{figure}
\noindent
With out loss of generality, we take $m=1$. Relevant component of the RG corrected momentum~\cite{Susskind2,Ageev:2018msv} corresponding to the coordinate $z(t)$ is:
\begin{equation}
 P_{z}=\frac{\dot{z}}{z\sqrt{\phi(z)}\sqrt{\bigg(1+\sqrt{1-4\lambda}\bigg)\phi(z)^2-\dot{z}^2}} \, .
\end{equation}
Using the mass shell constraint, the energy is:
\begin{equation}
E=-q\mu \left(1-\frac{z^2}{z_{+}^{2}} \right)+\frac{mN^2\phi(z)^{\frac{3}{2}}}{z\sqrt{\phi(z)^{2}N^{2}-\dot{z}^{2}}}
\label{energy}
\end{equation}
When the energy is negative the particle oscillates between two turning points $z_{*,\pm}$, with charge $q$($>0$) greater than the critical value:
\begin{equation}
q_{crit}=\frac{mN\sqrt{\phi(z_{*,-})}}{z_{*}A_{t}(z_{*,-})}\, .
\label{criticalcharge}
\end{equation}
\begin{figure}[h]
\centering
\begin{tabular}{c}
\includegraphics[width=.65\textwidth]{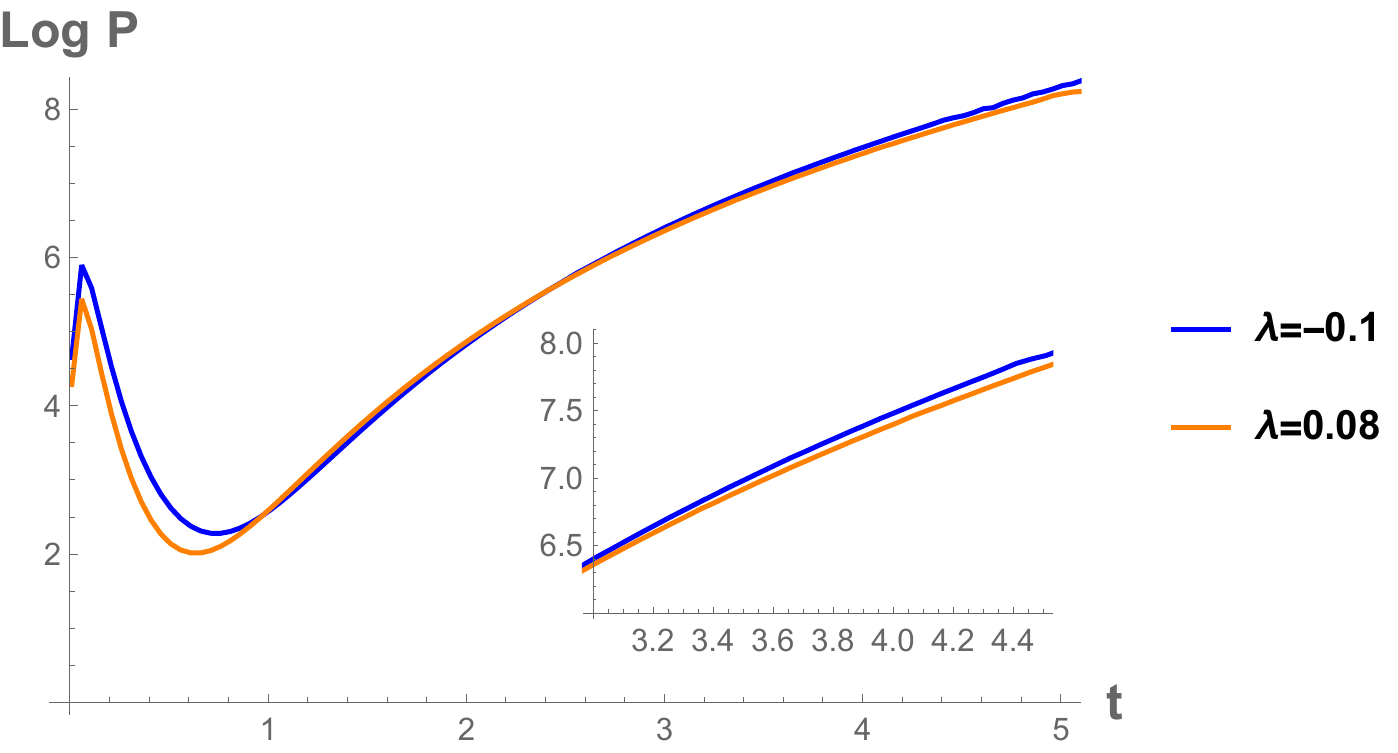}
\end{tabular}
\caption{Exponential growth of particle momentum for $q=0, l=1, z_h =1, \mu = 0.95 \mu_{extremal}$ for different values of the GB coupling constant.}
\label{PvstGB}
\end{figure}
\begin{figure}[h]
\centering
\begin{tabular}{c}
\includegraphics[width=.65\textwidth]{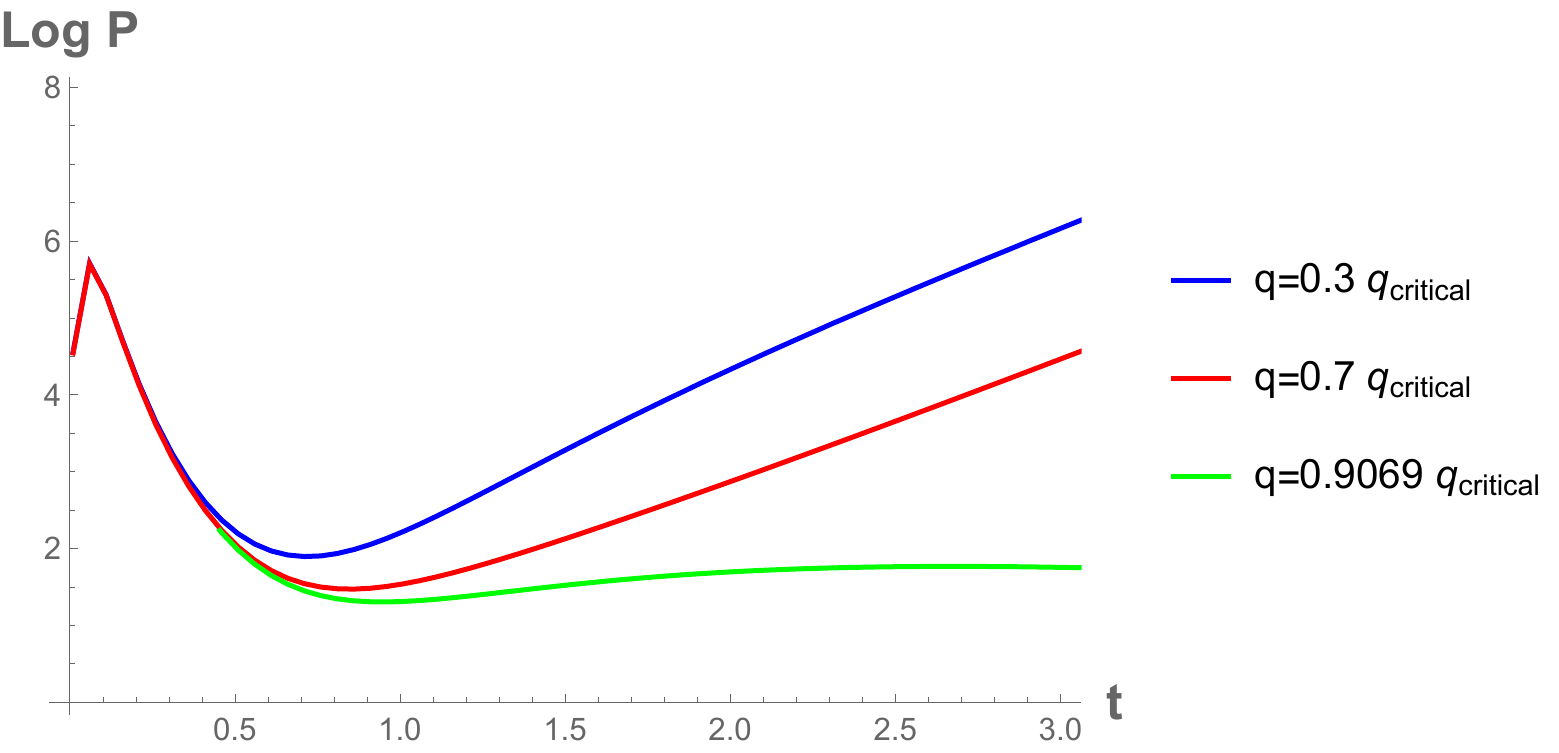}
\end{tabular}
\caption{Exponential growth of particle momentum shown for different values of charge. Parameters values used are: $l=1, z_h =1, \mu = 0.75 \mu_{extremal}, \lambda=0.08$. }
\label{PvstGBq}
\end{figure}
There are three regions of the charged black holes in AdS space-time in the near extremal limit: the Rindler region (close to the horizon $z_h=1$), a long throat region  and the boundary outer region (Newtonian)~\cite{Susskind2}. Particles start in the bulk, accelerate as they fall through the Newtonian barrier and fall in to the black hole asymptotically~\cite{Susskind2}. As noted in~\cite{Ageev:2018msv}, in the context of Reissner-Nordstrom black holes and even in the present case of GB black holes,  particles with different charges and bulk starting points $z_*$, probe different regions with characteristic trajectories\footnote{See figure-1 of~\cite{Ageev:2018msv} for a discussion of various trajectories of charged particles in Reissner-Nordstrom black holes, which are valid for any charged black hole close to extremality.}.  Depending on the charge value, some particles oscillate in the Newtonian and/or throat region and others fall in to the black hole asymptotically~\cite{Susskind2,Ageev:2018msv}.  The later set of orbits, where the particle is confined to the Rindler region with exponentially growing momentum at late times, happens universally for all neutral particles (in addition to charged particles with $q <  q_{crit}$) and corresponds to chaotic behaviour. To obtain the time evolution of particle momentum for different values of GB parameter in near extreme black holes, it is required to fix the chemical potential to be below the extremal value $\mu_{extremal}$ (given in eqn.(\ref{muext})). The $\mu_{extremal}$ is not a constant and varies with the GB parameter $\lambda$, as seen in figure-\ref{muextremal}. Thus, to ensure that the black hole is near extremal, while performing the numerical computations, we fixed the value of chemical potential to be $0.95 \mu_{extremal}$ and take the initial position of the particle to be at $z_*=0.1$ (for convenience of numerical computation), recursively at all stages. With this, the behaviour of the particle orbits and associated momentum evolution can be computed. From figure-\ref{PvstGB} it is clearly evident that the late time evolution of the neutral particle momentum is suppressed with the increase in GB coupling parameter $\lambda$. Thus the increase in GB coupling $\lambda$ suppresses chaos. Further, from figure-\ref{PvstGBq}, we conclude that chaos is also suppressed as the particle charge approaches the critical value $q_{critical}$. This might be a common feature in any general theory of gravity, supporting the observations in~\cite{Ageev:2018msv}. 

\subsection{Lifschitz and hyperscaling violating black holes} \label{HV}

Another class of very interesting models to further explore the size-momentum correspondence is those supporting both anisotropic and hyperscaling violating exponents. The holographic gravity set up for exploring these models is the Einstein-Maxwell dilaton system (see for instance \cite{Dong:2012se,Alishahiha:2012qu,Salvio:2013jia}). Holographic complexity in these models is studied for example in~\cite{Swingle:2017zcd,An:2018xhv,Alishahiha:2018tep}. Analytic solution for charged Lifshitz-hyperscaling violating black brane solution in four dimensions is known, and we start with the metric suitable for the present case given as~\cite{Kuang:2015mlf,Swingle:2017zcd,An:2018xhv,Alishahiha:2018tep}:
\begin{eqnarray} \label{hvads}
&&ds^2=u^{\theta-2}\Bigg(-u^{-2z+2}f(u)dt^2+\frac{du^2}{f(u)}+(dx^2+dy^2)\Bigg)\, . 
\end{eqnarray}
Models based on the above line element under consideration, exhibit a non-trivial scaling symmetry as,
\be
u\rightarrow \omega  u,\quad t\rightarrow \omega^z t,\quad  x\rightarrow \omega x, \quad y\rightarrow \omega y,
\quad ds \rightarrow \omega^{\frac{\theta}{2}} ds,
\ee
where the exponents $z$ and $\theta$ stand for anisotropic (Lifshitz) scaling and hyperscaling violating parameters.  As distances are no more invariant under scaling transformations, they indicate possible violations of hyperscaling in the boundary field theories. The lapse function appearing in eqn. (\ref{hvads}) is:
\begin{eqnarray}
&&f(u)=1-\Big(\frac{u}{u_{h}}\Big)^{2+z-\theta}+Q^2u^{2(z-\theta+1)}[1-\Big(\frac{u}{u_{h}}\Big)^{\theta-z}] \, ,
\end{eqnarray}
with $u_{h}$ as the horizon radius, and there are two field strengths,
\begin{eqnarray}
&&\mathcal{F}_{ut}=\sqrt{2(z-1)(2+z-\theta)}e^{\frac{2-\theta/2}{\sqrt{2(2-\theta)(z-1-\theta/2)}}\phi_{0}}u^{-(1+z-\theta)} \, , \\
&&F_{ut}=Q\sqrt{2(2-\theta)(z-\theta)}e^{-\sqrt{\frac{z-1+\theta/2}{2(2-\theta)}}\phi_{0}}u^{(z-\theta+1)} \, ,\\
&&e^{\phi}=\frac{e^{\phi_{0}}}{u}\sqrt{2(2-\theta)(z-1-\theta/2)}\, .
\end{eqnarray}
The field strength $F_{ut}$ gives rise to the gauge field $A_t$ below, that contributes to the charged solution. $\mathcal{F}_{ut}$ on the other hand, together with dilation $\phi$ creates a anisotropic scaling.  We can write the gauge fields as
\begin{eqnarray}
&&\mathcal{A}_{t}=-\cancel{\mu}\Bigg[1-\Big(\frac{u_{h}}{u}\Big)^{2+z-\theta}\Bigg]~\text{with}~\cancel{\mu}=\sqrt{\frac{2(z-1)}{2+z-\theta}}e^{\frac{2-\theta/2}{\sqrt{2(2-\theta)(z-1-\theta/2)}}\phi_{0}}u_{h}^{(\theta-2-z)} \, ,\nonumber\\
&&A_{t}=\mu\Bigg[1-\Big(\frac{u}{u_{h}}\Big)^{z-\theta}\Bigg]~~\text{with}~~\mu=Q\sqrt{\frac{2(2-\theta)}{z-\theta}}e^{\sqrt{\frac{z-1+\frac{\theta}{2}}{2(2-\theta)}}\phi_{0}}u_{h}^{z-\theta} \, .
\end{eqnarray}
Hawking temperature of the black hole in this case is 
\begin{equation}
T=\frac{(2+z-\theta)u_{h}^{-z}}{4\pi}\Bigg[1-\frac{(z-\theta)Q^{2}}{2+z-\theta}u_{h}^{-2(\theta-z-1)}\Bigg] \, .
\end{equation}
The consistency conditions for the existence of solution suggest the following constraints~\cite{Kuang:2015mlf,Swingle:2017zcd,An:2018xhv,Alishahiha:2018tep}:
\begin{eqnarray} \label{range}
&&0\leq\theta\leq2(z-1)~~\text{for}~~1\leq z<2 \, ,\nonumber\\
&&0\leq\theta< 2 ~~\text{for}~~z\geq2 \, .
\end{eqnarray}
In the extremal limit, the Hawking temperature is zero and one gets
\begin{eqnarray}
Q_{extremal}=\sqrt{\frac{2+z-\theta}{z-\theta}}u_{h}^{(\theta-1-z)} \, ,\nonumber\\
\mu_{extremal}=\frac{\sqrt{2(2-\theta)(2+z-\theta)}}{z-\theta}\frac{1}{u_{h}} \, .
\end{eqnarray}
Following the methodology in last section, the critical charge and RG corrected particle momentum can be computed to be:
\begin{eqnarray} \label{Pzhv}
q_{crit}&=&\frac{\sqrt{f(u_{*,-})}u_{*,-}^{\frac{\theta}{2}-z}}{\mu(1-(\frac{u}{u_{h}})^{z-\theta})}.\nonumber\\
P_{z}&=&\frac{\dot{u}}{\sqrt{u^{2-\theta}f(u)(u^{2-2z}f(u)^2-\dot{u}^{2})}} \, .
\end{eqnarray}
\begin{figure}[h]
\centering
\begin{tabular}{c}
\includegraphics[width=.55\textwidth]{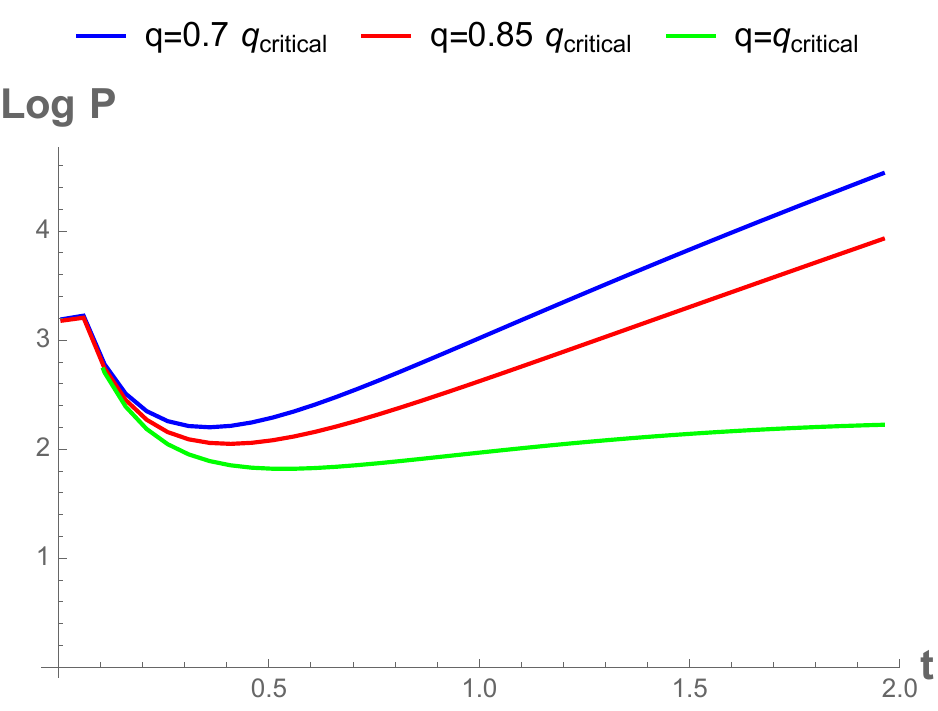}
\end{tabular}
\caption{Exponential growth of particle momentum in Hyperscaling violating theories shown for different values of charge. Parameters values used are: $z =1.5, \theta=0.5, \mu = 0.75 \mu_{extremal}$. }
\label{HVPtq}
\end{figure}

\begin{figure}
	\begin{center}
		{\centering
			\subfloat[]{\includegraphics[width=6cm,height=5cm]{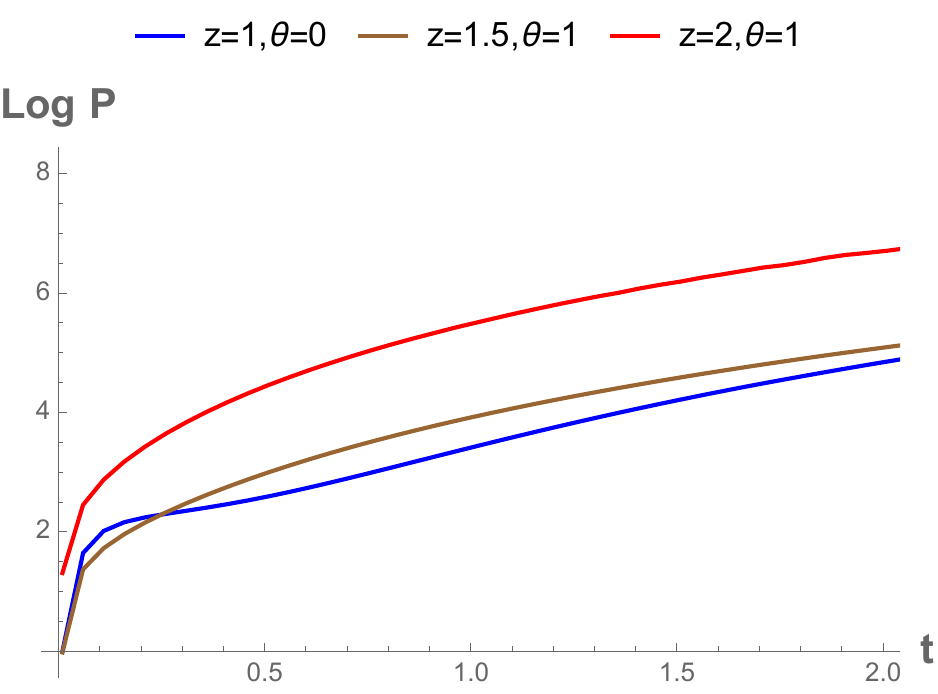} } \hspace {1.5cm}
						\subfloat[]{\includegraphics[width=6cm,height=5cm]{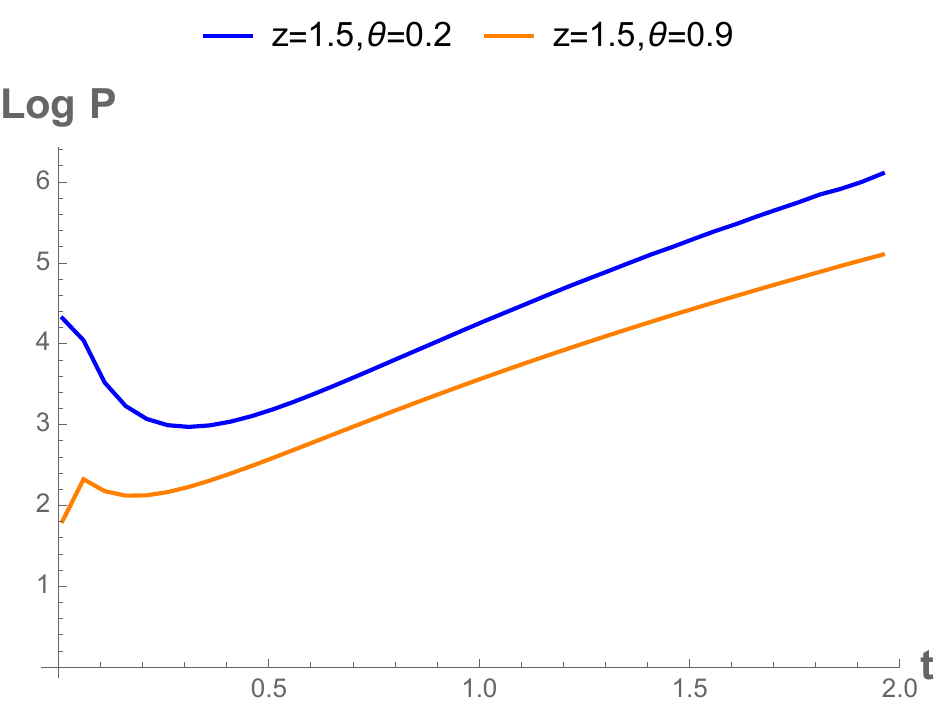} } 
			\caption{Time evolution of particle momentum in Hyperscaling violating theories for $q=0, \mu = 0.95 \mu_{extremal}$: Suppression in chaotic behaviour for different cases. The $z=1,\theta=0$ case corresponds to the result in Reissner-Nordstrom black holes:  (a) variation w.r.t. z (b) variation w.r.t. $\theta$}  \label{HVPt}
		}
	\end{center}	
\end{figure}
As in the previous section, one can obtain the late time behaviour of particle momentum given in eqn. (\ref{Pzhv}) numerically. The results are plotted in figures \ref{HVPtq} and \ref{HVPt}, which show the late time evolution of particle momentum for charged and neutral cases, respectively. For charged particles, the late time momentum behaviour in figure-\ref{HVPtq}) shows that the chaotic nature is suppressed as particle charge $q$ approaches $q_{crit}$, and stops evolving when $q=q_{critical}$. This is consistent with the observations in the previous section and seems to be a general feature. For neutral particles, the chaotic behaviour increases with increasing $z$ (at fixed $\theta$) or decreasing $\theta$ (at fixed $z$). In the case $z=1,\theta=0$, the metric in eqn. (\ref{hvads}) goes back to that of a planar Reissner-Nordstrom black hole and  the results are identical to the behaviour of bulk momentum studied in~\cite{Ageev:2018msv}. Thus, as compared to the case $z=1,\theta=0$ case, the bulk momentum at late times is always higher for any value of anisotropy (say, $z >1$) and hyperscaling violating parameter (with in the range specified in eqn. (\ref{range}). Thus, as per the proposals in~\cite{Susskind1,Susskind:2019ddc,Susskind2} and  eqn. (\ref{triple}), the growth rate of complexity is also expected to be higher in anisotropic theories.   The growth rate of complexity was studied using the holographic CA and also the CV proposals in~\cite{Swingle:2017zcd,An:2018xhv,Alishahiha:2018tep}. The enhancement in complexity for $z >1$ as compared to $z=1$ was already noted through direct computation in~\cite{Swingle:2017zcd,An:2018xhv,Alishahiha:2018tep}.

\section{Growth of operator size on the boundary} \label{size}

The growth of operator size on the boundary can be computed by considering the bulk dynamics of a particle in the bulk geometry. The wavelength associated to the gluon excitations is given as $\lambda=\frac{2\pi l}{r}$, $r$ being the radial location of the bulk particle. The inverse dependence of wavelength on $r$ is due to the UV-IR connection \cite{Susskind:1998dq}. The size of the operator $(s(r))$ which is nothing but the number of gluon excitations is then given as,
\begin{equation}\label{sr}
s(r)=El\lambda=s_{0}\frac{r_{c}}{r},
\end{equation}
where, $s_{0}=\frac{2\pi E l^2}{r_{c}}$ with $r_{c}$ as some radial cut-off.
Solving the geodesic equation of a bulk particle propagating along the radial direction in empty $AdS$ background and also using the above equation (\ref{sr}), the time dependence of the size of boundary operator $(s(t))$ has been computed in \cite{Susskind:2020gnl} with the observation that it grows linearly with time, namely,
\begin{equation}
s(t)=2\pi E t.
\end{equation}
The same linear behaviour was also observed for a bulk particle moving in a three dimensional BTZ black hole but only at time scale satisfying $t<\frac{l^2}{r_{h}}$, with $r_{h}$ as the black hole horizon.

The geodesic equation of a particle with four velocity $v^{\alpha}$ and proper time $\tau$ is given as,
\begin{equation}
\frac{dv^{\alpha}}{d\tau}=-\Gamma^{\alpha}_{\beta\gamma}v^{\beta}v^{\gamma}.
\end{equation}
Here we redefine the radial coordinate as $u=r_{h}/r$ and assume that the bulk particle starts from a cut-off position $u_{c}=r_{h}/r_{c}$ along the radial direction. Solving the above geodesic equation along with the relation for the line element $g_{\mu\nu}\frac{dx^{\mu}}{d\tau}\frac{dx^{\nu}}{d\tau}=-1$, the dynamics of the bulk particle can be determined. The rate of operator growth then follows from equation (\ref{sr}). In this paper we have solved the geodesic equation numerically for the radial dynamics of a neutral bulk particle moving in the charged Gauss-Bonnet black hole background and also in Lifshitz-Hyperscaling violating theories in AdS geometry to obtain the operator size growth as a function of time.
\begin{figure}[h]
\centering
\begin{tabular}{c}
\includegraphics[width=.55\textwidth]{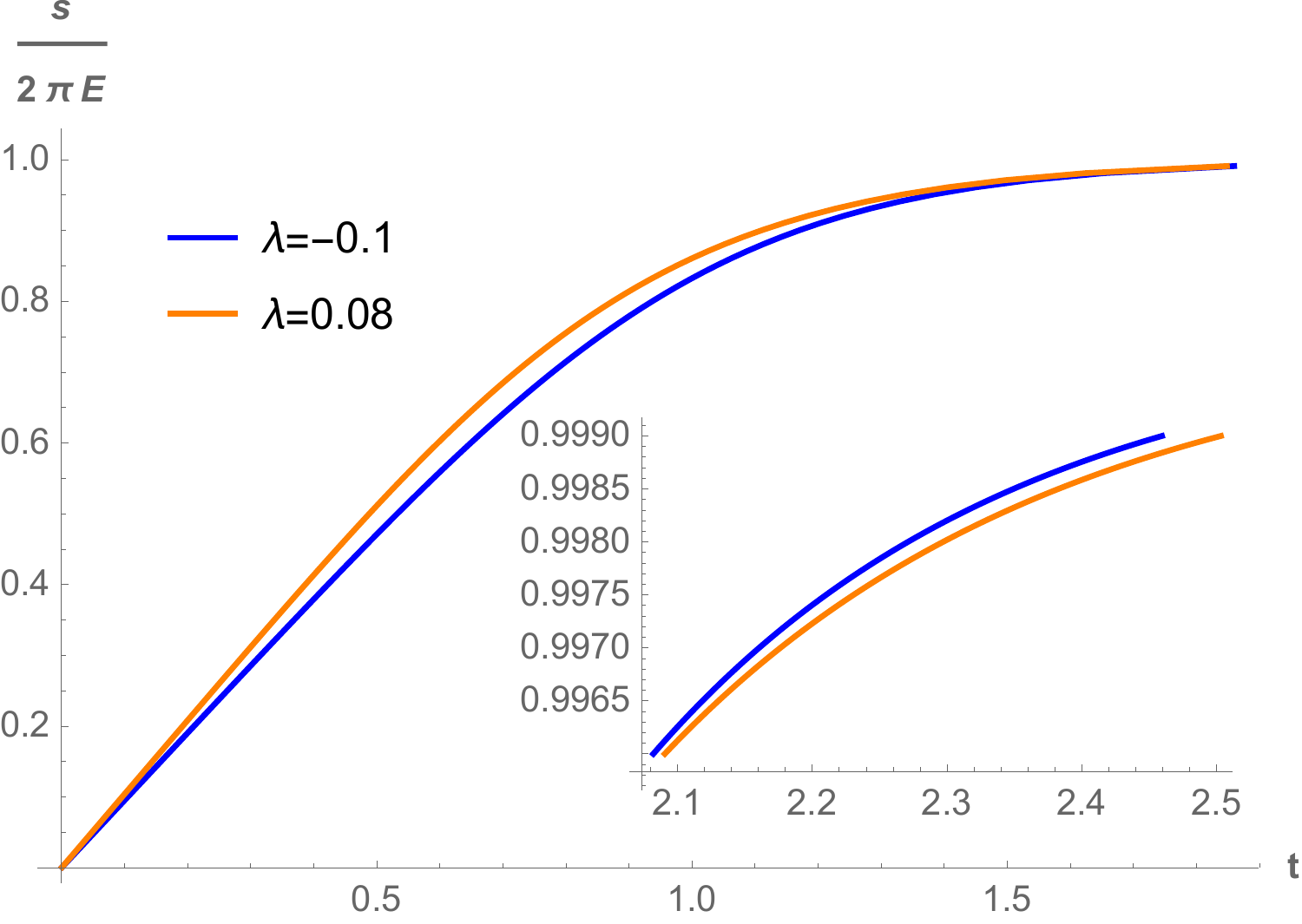}
\end{tabular}
\caption{Plot showing the nature of the operator size growth with increasing time for different values of Gauss-bonnet coupling for $Q=0.5,r_h=1,l=1$.}
\label{11}
\end{figure}
In Figure-{\bf 6}, we have shown the nature of the operator size growth with increasing time for the charged Gauss-Bonnet black hole background with two different values of the Gauss-Bonnet coupling. Let us mention the following observations regarding the operator size growth as shown in Figure-{\bf 6}.
\begin{itemize}
\item Note that the operator size grows linearly at early times for the two different values of coupling. Similar kind of linear growth has been obtained in \cite{Susskind:2020gnl} for empty $AdS$ and also for three dimensional BTZ black hole background in the time regime $0<t<\frac{l^2}{r_{h}}$ (in this paper we set $l=r_{h}=1$ for all the calculations).
\item Another interesting observation is that the rate of operator size growth increases with the Gauss-Bonnet coupling which is evident from Figure-{\bf 6} as the orange curve which corresponds to the value of coupling $\lambda=0.08$ has a greater slope than the blue curve with $\lambda=-0.1$. However at late time we observe a completely opposite behaviour for the operator growth (see the inset plot). In particular, the growth of the operator size decreases as the value of the coupling is increased indicating a suppression in the chaotic behaviour of the system. In the previous section we have drawn a similar conclusion by computing the linear momentum of the bulk particle which slows down with increasing values of the coupling. Regarding the late time growth of complexity for the Gauss-Bonnet background, it is shown in \cite{An:2018dbz} that the complexity growth rate also decreases with increasing coupling. Hence the above result is in accordance with the connection between momentum, size and complexity as conjectured in \cite{Susskind:2020gnl}.
    \item Finally, we observe that at late time the operator growth saturates which has been shown previously for the black hole system in \cite{Roberts:2014isa}.
\end{itemize}
\begin{figure}[h]
\centering
\begin{tabular}{ccc}
\includegraphics[width=.45\textwidth]{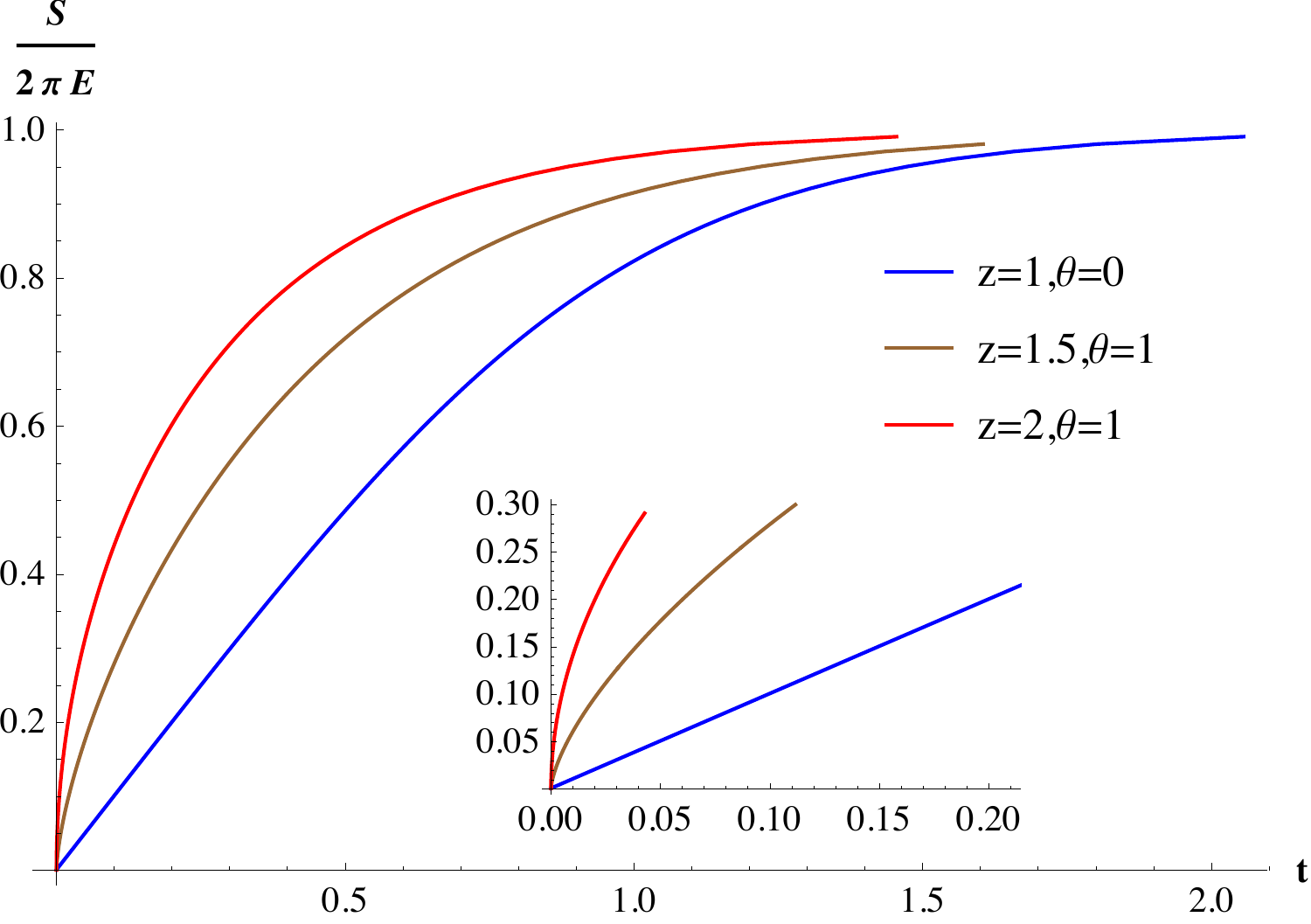}&&
\includegraphics[width=.45\textwidth]{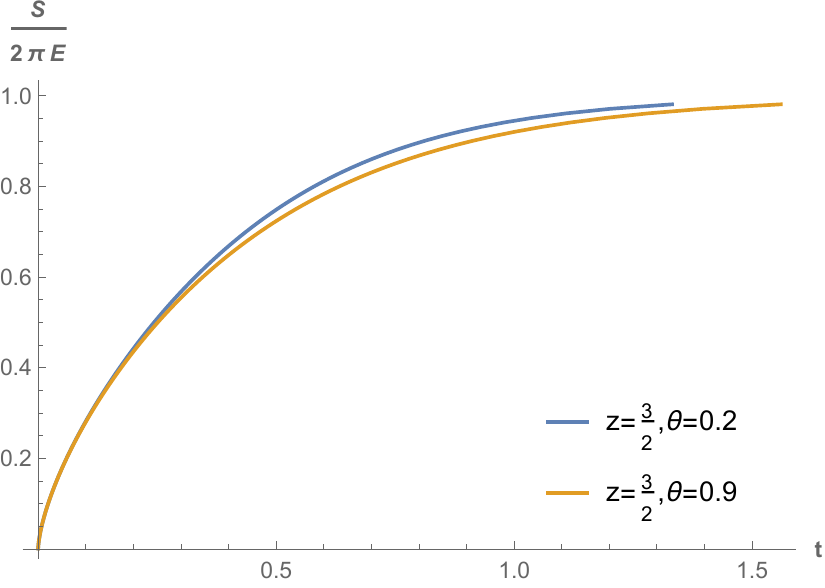}\\
a)&&b)
\end{tabular}
\caption{Plots showing the operator size growth for the Lifshitz background with different values of $z$ and $\theta$ for $Q=0.1,r_h=1$.}
\label{12}
\end{figure}
Figure-{\bf 7} captures the operator size growth with time for black hole background with Lifshitz scaling. Here we have considered different values of the exponents $z$,$\theta$ to the corresponding modifications to the rate of operator growth. In the following we lists some of the interesting observations regarding the plots in Figure-{\bf 5}.
\begin{itemize}
\item In Figure-{\bf 7a}, we see that the operator growth rate increases with $z$ for a fixed value of $\theta=1$. For instance the brown coloured curve with $z=1.5$ always stays below the red one which corresponds to the value $z=2$. Similar kind of variation has been observed in the earlier section regarding the momentum computation in this background.
    \item Also note from the same figure (see the inset plot) that the linear growth at early time is obtained with $z=1, \theta=0$. However as the values of $z$ and $\theta$ is increased, the linear dependence shows up only at very small time scale. Moreover, for fixed $\theta$, the operator growth saturates at smaller time as the value of $z$ is increased.
        \item In Figure-{\bf 7b}, we have shown the operator growth rate for a particular value of $z=1.5$ but with two different $\theta=0.2,0.9$. In this case, the operator growth rate decreases with $\theta$ which is very similar to the momentum growth of a bulk particle obtained in previous section.
\end{itemize}

\section{Remarks} \label{conclude}

In this paper, we studied the late time behaviour of radial momentum of particles falling into black holes in various models such as Gauss-Bonnet and Lifshitz-Hyperscaling violating theories and matched its features with the operator growth on the boundary. We closely followed the methods in~\cite{Ageev:2018msv}, where the behaviour of charged particles falling into near extremal charged black holes in AdS was investigated and found that, ``charge stops the fall". The motivation of this paper was to see how the natural parameters of the model (such as $\lambda$ in Gauss-Bonnet theory and $z,\theta$ in Lifshitz theories), effect the chaotic aspects of the theory and hence neutral particles were chosen as probes for most part. Our analysis relied on the holographic proposal in~\cite{Susskind1,Susskind:2019ddc,Susskind2}. In the Gauss-Bonnet theory, we observed a suppression in chaotic behaviour as the coupling constant $\lambda$ is increased. This feature is consistent both with the late time analysis of particle momentum in the bulk as well as the operator growth in the boundary, giving support to the first two relations in equation-(\ref{triple}). In fact, our results are also in conformity with the last relation in equation-(\ref{triple}), since, independent calculations on the growth rate of complexity in GB black holes was studied in~\cite{An:2018dbz} and, the suppression in the rate of complexity with increase of GB coupling $\lambda$ was noted. The conclusions in~\cite{An:2018dbz} were based on the complexity-volume (CV) proposal\cite{Stanford:2014jda,Alishahiha:2015rta}, although a comparison with complexity-action (CA) proposal~\cite{Brown:2015bva,Brown:2015lvg,Lehner:2016vdi} was also made. The suppression in complexity was observed to be more universal in the CV proposal, than in the CA proposal.\\

\noindent 
Some understanding of our results follows from the holographic analysis of quantum information scrambling, where it is believed that the black hole systems scramble information extremely fast \cite{Sekino:2008he}. However, it is shown explicitly in \cite{Shenker:2014cwa} that any stringy correction (Gauss-bonnet correction) included to the Einstein gravity reduces the rate of information scrambling and consequently the time required for the black hole to thermalise increases. As a concrete example to the above discussion one can consider the Gauss-Bonnet correction term in the supergravity action. The Gauss-Bonnet coupling being proportional to the string parameter $\alpha^{\prime}$ (bosonic and heterotic string theory) can tune the coupling constant of the dual field theory. Hence as one increases the Gauss-Bonnet coupling, the corresponding dual field theory becomes weakly coupled \cite{Brown:2015lvg} and it requires longer time for the information to spread across the entire system. A quantitative manifestation of quantum information scrambling is the growth of the boundary operator size over time. So it is expected that operator growth will be suppressed due to the finite Gauss-Bonnet coupling and is consistent with what was found in this paper by direct computation.\\

\noindent 
For  Lifshitz-hyperscaling violating theories, the particle momentum as well as the operator growth rate decreases with $\theta$ and increases with $z$. In particular, the operator size is found to be higher as compared to the $z=1,\theta=0$ case (corresponding to particle falling in to Reissner-Nordstrom black holes) and this is in conformity with the results on complexity growth rate studied earlier in these models~\cite{Swingle:2017zcd,An:2018xhv,Alishahiha:2018tep}. We also considered charged particles falling into black holes, to see whether the conclusions of~\cite{Ageev:2018msv} are valid in general theories of gravity. We found that charge does stop the fall, even in general theories of gravity generalising the results of~\cite{Ageev:2018msv}]. Figures 3 and 4 show that as the charge of the particle increases, chaos is suppressed. Thus, the results on late time behaviour of particle momentum and hence the operator size growth studied in this paper, are found to be consistent with the known results on growth of holographic complexity in these models, justifying the proposals in~\cite{Susskind1,Susskind2,Susskind:2019ddc}. It would be interesting to extend the results of this work by doing more elaborate quantitative computations. For instance, the evolution of CFT under a local quench is well studied\cite{Calabrese:2007mtj} and corresponds to a deformation of the black hole perturbed by the particle falling on the horizon~\cite{Nozaki:2013wia}. The behaviour of complexity in such a scenario has been studied in~\cite{Ageev:2019fxn} and hence it is tempting to continue by studying both the size as well momentum behaviours by taking care of such back reaction effects. Another possible avenue is to explore in general theories of gravity, is the PC correspondence for systems with finite size. Here, although the operator growth supposedly stops at scrambling time, the complexity continues to grow at much larger times and involves trajectories in black hole interior~\cite{Barbon:2019tuq}.

\section*{Acknowledgements}
C.B. would like to thank the DST (SERB), Government of India, for the Mathematical Research Impact Centric Support (MATRICS) grant no. MTR/2020/000135. We are grateful to the anonymous referee for a detailed and thought provoking report which has improved the paper.



\end{document}